# Structural, dielectric, and electrical transport properties of $Al^{3+}$ substituted nanocrystalline Ni-Cu spinel ferrites prepared through the sol-gel route

M. M. Rahman[a], N. Hasan[b], M. A. Hoque[c], M. B. Hossen[d], M. Arifuzzaman[e,*]

[a]Department of Industrial and Production Engineering, Bangladesh University of Textiles, Dhaka-1208, Bangladesh
[b]Department of Electrical and Computer Engineering, North South University, Dhaka-1229, Bangladesh
[c]Bangladesh Council of Scientific and Industrial Research, Dhaka-1205, Bangladesh
[d]Department of Physics, Chittagong University of Engineering and Technology, Chattogram-4349, Bangladesh
[e]Department of Mathematics and Physics, North South University, Dhaka-1229, Bangladesh

*Corresponding author:

M. Arifuzzaman (md.arifuzaman01@northsouth.edu)

**Abstract**

In this study, a series of nanocrystalline ferrites of $Ni_{0.7}Cu_{0.30}Al_xFe_{2-x}O_4$ (x=0.00 to 0.10 with a step of 0.02) has been synthesized through the sol-gel auto combustion technique. The structural, morphological, dielectric, and electrical properties of the Ni-Cu spinel ferrite nanoparticles are analyzed due to the substitution of $Al^{3+}$ content. The crystalline and structural characteristics of the prepared nanoparticles (NPs) have been studied employing the x-ray diffraction (XRD) spectra and FTIR analysis. The extracted XRD patterns assure the single-phase cubic spinel structure of all samples with homogeneity and no impurity, which indicates the yielding of high crystalline NPs. The average crystallite size of the synthesized ferrite nanoparticles is found in the range (55.63–70.74 nm) and the average grain size varies from 59.00 to 65.00 nm. FTIR study also confirms the formation of spinel structures in the prepared Ni-Cu ferrite nanoparticles. A slight decrease of average grain size with increment of $Al^{3+}$ content is observed in the surface morphological study carried out by the field emission scanning electron microscopy (FESEM). The studied materials are found in semi-spherical shapes, showing the multi-domain grains separated by grain boundaries with some agglomerations. The chemical composition study for the synthesized NI-Cu spinel ferrites using energy dispersive x-ray (EDX) ensures the presence of each component in appropriate proportions in each sample. The dielectric dispersion nature of all investigated materials is reflected in the current study up to the frequency of 10 kHz. Because of the high resistive grains, the value of ′ is higher at low frequencies, resulting in space charge polarization. However, the effect of cation distributions on A and B sites on the grain dependent space charge polarization nature is reflected in the dielectric constant. The sample with x = 0.1 demonstrates that the space charge polarization has increased, resulting in a higher dielectric constant value. The impedance spectroscopy confirms the non-Debye relaxation phenomena of the synthesized nanomaterials. The contribution of grains and grain boundaries is resolved through the modulus study of the materials, which reconfirms their dielectric relaxation. The trend in variation of AC resistivity suggests the normal behavior of the materials with varying frequencies, which is explained by the hopping mechanism.

1. **Introduction:**

Magnetic ferrite nanoparticles in the spinel phase have been considered as the influential class of materials, which are employed in various high-frequency device applications[1]. The cubic spinel structure has the chemical formula of $AB_2X_4$, where the anions X are occupied by O atom as of metal oxides forming the cubic close-packed lattice, tetrahedral interstices fill the A site as the 'network formers' and octahedral interstices occupy the B site as the 'modifiers', called the Ferro-spinel and semiconductor in nature [2–5]. Most spinel ferrites belong to the space group of *Fd3m* (No. 227, Z = 8), which provide the highest symmetrical face-centered cubic (FCC) spinel structure. A spinel unit supercell's crystal is formed by 8 A-sites and 16 B-sites cations. Based on the distribution of divalent metal ions and trivalent ferric ions over A and B sites, spinel ferrites are of three classes; normal spinel, inverse spinel, and mixed spinel [6].

Magnetically soft spinel ferrites are used in a large spectrum of biomedical and industrial applications including medical treatments, such as magnetic resonance imaging, antenna fabrication, computer memories, energy storage in supercapacitors, high-density information storage, high-frequency transformers, hyperthermia treatment, multi-layered chip inductors, water purification methods, sensing of nucleic acid, separation of DNA and RNA, gene therapy and delivery, ferrofluids and so on [7–13]. The advancement of electronic devices is now moved to integrated circuits-based technology, where highly efficient transistors are increasing gradually in accordance with Moore's law which requires nano-level engineering and fabrication. Thus, in contrast to bulk materials, researchers are now focusing on nanocrystalline ferrites' for utilizing them in the advancement of nano-technological devices. The physical and chemical characteristics of ferrite nanomaterials mostly depend on their scale size, shape, or morphology. The structural parameters such as crystal size and lattice parameters are somehow linked to the electrical and magnetic properties of ferrite nanoparticles. Therefore, the controlling of several factors such as the particle size, surface-to-volume ratio, magnetic anisotropy eventually improves the electronic properties of magnetic nanoparticles in the spinel phase, owing to their transitions from bulk to nano-shape.

Researchers are continuously paying their efforts to employ an easy and efficient method for yielding the nanocrystalline ferrites to tailor their structural, dielectric, electric, and magnetic properties under favorable environmental conditions. Various techniques have been deployed to synthesize nanostructured ferrite materials till now *viz.* sol-gel auto combustion, co-precipitation, high-energy milling, hydrothermal synthesis, precursor method, mechanochemical route, and microwave hydrothermal [7–9,14–17]. Among these, the sol-gel route appears to be a prominent method for preparing ferrite nanoparticles, as it is eco-friendly, less expensive, and effective without the involvement of expensive equipment to maintain a good stoichiometry during the synthesis process. The sol-gel is a wet chemical method, which is widely used due to its potential advantages such as enhanced control over homogeneity, elemental composition, and powder morphology with a uniform narrow particle size distribution at relatively low temperature [7,16–18].

Researchers attempted sporadically to study the structural, electrical, morphological, photocatalytic, and magnetodielectric properties of Ni-Cu series ferrite NPs demonstrating various effects of doping on the properties of the materials crystal. Doping is an effective method to ameliorate the applications to a broad range by achieving excellent optoelectronic properties. Investigations are still continued with selecting different atoms as dopants or substitutions in A and B- sites to tailor the physical, structural and electromagnetic properties of Ni-based mixed spinel ferrite nanoparticles [19–28]. *Munir et al.*[29] conducted an experiment with a noble nanocomposite $CuFe_2O_4/Bi_2O_3$ by introducing $Bi_2O_3$ nano-petals into the porous $CuFe_2O_4$ and observed a significant increase in the photocatalytic activity in effect of photo-degradation activity. However, the investigated nanocomposite has remarked with an excellent magnetic separation at room temperature for the reduced recombination and improved separation of electron-hole pairs. Carbon coated highly active magnetically recyclable hollow nano-catalysts have been synthesized by *Shokouhimehr et al.*[30], where the authors projected that the prepared nanocomposite can be used as a general platform for loading other noble metal catalyst nanoparticles, resulting in high yields (up to 99 percent) in selective nitroarenes reduction and Suzuki cross-coupling reactions. Furthermore, magnetic properties revealed that the catalysts could be easily separated using a suitable magnetic field and recycled five times in a row. Moreover, *Rahman et al.*[31] thoroughly investigated the photocatalytic efficiency and recycling stability of rGO supported cerium substituted nickel ferrite nanoparticles under visible light illumination. According to their findings, $NiCe_yFe_{2-y}O_4$/rGO (NCFOG) nanocomposite outperformed $NiCe_yFe_{2-y}O_4$ nanoparticles by two times in photocatalytic efficiency and recycling stability, which is attributed to the formation of NCFOG heterojunction that enables in the separation of photo-induced charge carriers while maintaining a strong redox ability. Recently, *M. Arifuzzaman et al.*[17] studied Cu substituted Ni-Cd ferrite NPs and reported the decrease of average crystallite size and saturation magnetization of $Ni_{0.7-x}Cu_xCd_{0.3}Fe_2O_4$ up to x=0.2. Besides, *V. A. Bharati et al.*[32] reported the influence of parallel doping of $Al^{3+}$ and $Cr^{3+}$ on the structural, morphological, magnetic, and MÖssbauer properties of Ni ferrite NPs and justified their suitability in HF device applications. In [33], *K. Bashir et al.* revealed the electrical and dielectric properties of Ni-Cu ferrite NPs with the doping of $Cr^{3+}$, making them the potential for HF applications and photocatalytic activity. *Le-Zhong Li et al.*[34] examined the $Al^{3+}$ substituted Ni-Zn-Co ferrites and observed a decrease in saturation magnetization at >0.10. They reported about the metal-semiconductor transition behavior of Ni-Zn-Co ferrites as an effect of varying temperature and the increase of dc resistivity with Al content was found. The structural and magneto-optical properties of Ni ferrite NPs were propounded in [35], where the authors calculated the electronic bandgap of 1.5 eV and observed a decrease in saturation magnetization and $T_c$ with $Al^{3+}$ content. In addition, density functional theory (DFT) based simulation was employed in estimating the electronic structure of CuO NPs with the optimized geometric crystal calculation, which showed the variation of energy band gap with Al content in the samples [36]. The effect of doping materials on the characteristics of the different spinel ferrite nanoparticles are also available in the literature, Zn ferrite [37,38], Ga ferrite [39], Co ferrite [40–42], Fe ferrite [43], Mg ferrite [25,44], and Ni-Zn ferrite [45,46].

However, as per literature survey, no study has been found yet on the structural, dielectric, and electrical properties of $Al^{3+}$ substituted nanocrystalline Ni-Cu spinel ferrites. Therefore, it is important to perceive the role of Al substitution on nano-crystallinity and the physical characteristics of Ni-Cu ferrite NPs. Henceforth, the present study aims to explore the influence of $Al^{3+}$ substitution on the structural, dielectric dispersion and electrical conductivity properties of the synthesized nanocrystalline $Ni_{0.70}Cu_{0.30}Al_xFe_{2-x}O_4$ (x=0.00 to 0.10 with a step of 0.02) ferrites through the sol-gel process.

## 2. Experimental details

*Materials:*

To synthesize the studied nanocrystalline Ni-Cu spinel ferrites, analytical-grade reagents-nickel (II) nitrate $Ni(NO_3)_2.6H_2O$ (98%), copper (II) nitrate $Cu(NO_3)_2.3H_2O$ (95-103%), ferric (III) nitrate $Fe(NO_3)_3.9H_2O$ (98%), and aluminum (III) nitrate $Al(NO_3)_3.9H_2O$ (98%) were used in this experiment which purchased from the Research-Lab Fine Chem.

### 2.1 Synthesis of Ni-Cu-Al nanoparticles:

Derivatives of $Ni_{0.70}Cu_{0.30}Al_xFe_{2-x}O_4$ ($0 \leq x \leq 0.1$) nanoparticles with a step of 0.02, were synthesized by the sol-gel process. In this process, metal materials of $Ni(NO_3)_2.6H_2O$, $Cu(NO_3)_2.3H_2O$, $Fe(NO_3)_3.9H_2O$, and $Al(NO_3)_3.9H_2O$ were taken as raw materials and dissolved them in ethanol and mixed in certain proportions with a magnetic steer to make a homogeneous solution. The pH of the mixture was kept at 7 using the liquid $NH_4OH$ solution and the sol was continued to heat up to a temperature of 70°C until turning it into a form of dry gel. In an electric oven, the dried gel was heated at 200°C for 5 hours, during which a self-ignition process occurred and the compositions gradually became fluffy-loose powder. To obtain the resulting ingredients in a highly crystalline form, the derived powder was annealed at 700°C for another 5 hours to eliminate any impurity present in the samples. The powder was further homogenized by grinding it in a hand-milling process in a mortar. A hydraulic press of 65 MPa was then applied to the samples for 2 minutes to condense and turned them into disk-shaped samples. The prepared samples were 12 mm in diameter and 2.3 mm in thickness. Powder samples were finally sent for further study on dielectric and electrical measurements.

## 2.2 Characterization and property measurements

The structural parameters of the yielded nanocrystalline ferrites were determined through the powder x-ray diffractometer (XRD) analysis using the model PW3040, with CuK$_\alpha$ radiation of $\lambda = 1.5418$Å. The lattice parameter, crystal size (D), and the displacement density were retrieved by using the XRD data. The theoretical density ($\rho_{th}$), micro-strain ($\varepsilon_{ms}$), lattice strain ($\varepsilon_{ls}$), and stacking faults in the crystal structure were also determined. The lattice parameter ($a$) and crystallite size (D) were measured by the following relations [20]:

$$a = d_{hkl}\sqrt{(h^2 + k^2 + l^2)} \qquad (1)$$

$$D = \frac{0.9\lambda}{\beta_{hk} \cos\theta} \qquad (2)$$

where, $\lambda$, $\beta_{hkl}$, $\theta$, and $d_{hkl}$, respectively, indicate the wavelength of the X-ray, the full width at half maximum (FWHM) at the most prominent peak (311), the Bragg's angle, and the distance between adjacent planes. Fourier transform infrared (FTIR) spectroscopy was performed to investigate the about spinel phase in structure in all of the prepared samples. The morphology of the studied materials has been interrogated by the Field Emission Scanning Electron Microscopy (FESEM) (JEOL-JSM 7600F model). The Wynne Kerr Impedance Analyzer (model:6500B) was used to determine the complex dielectric ($\varepsilon^*$), AC resistivity ($\rho_{AC}$), complex electric modulus (M*), and complex impedance (Z*) of nanocrystalline ferrite samples.

## 3. Result and Discussion:

### 3.1 Structural analysis:

XRD patterns of $Al^{3+}$ substituted Ni-Cu ferrites annealed at 700°C, are illustrated in Fig. 1, where the peaks are resulted due to diffractions from the planes of (111), (220), (311), (222), (400), (422), (511), and (440). The peaks are shaped well-defined with a homogeneous distribution of nanoparticles, which attest to their highly crystalline nature with no impurity. Such peaks indicate the cubic single-phase formations of the spinel materials [32–34]. The peak diffracted from the plane (311) is found as the high intensity, which was used to determine the average crystallite size of the materials (see Table 1) using Debye-Scherer's equation. The lattice constant ($a_0$) values are calculated by the Nelson-Riley technique and unit cell volumes (V) of the compositions are listed in Table 1. The decreasing trend of lattice constant and cell volume with increasing $Al^{3+}$ content is observed, which is due to the replacement of larger ionic (0.67 Å) cations by that with smaller radius (0.51 Å). As $Al^{3+}$ is replacing to the place of $Fe^{3+}$ in the investigated ferrites, the unit cell becomes shrinkage, as a result, both $a_0$ and V decrease linearly with $Al^{3+}$ content, well satisfied by Vegard's law [47,48]. As appeared in Table 1, the average crystallite size decreases with $Al^{3+}$ content, which might be because of the ionic radius difference between $Al^{3+}$ and $Fe^{3+}$ ions, offering redistribution of cations in A and B sites which ultimately cause the increase in stress and strain of the samples. Lattice spacing is determined by the following equation:

$$d = \frac{n\lambda}{2sin\theta} \quad (3)$$

where d is the inter-spacing distance between crystal planes and the value of n is taken as 1, which represents the order of diffraction.

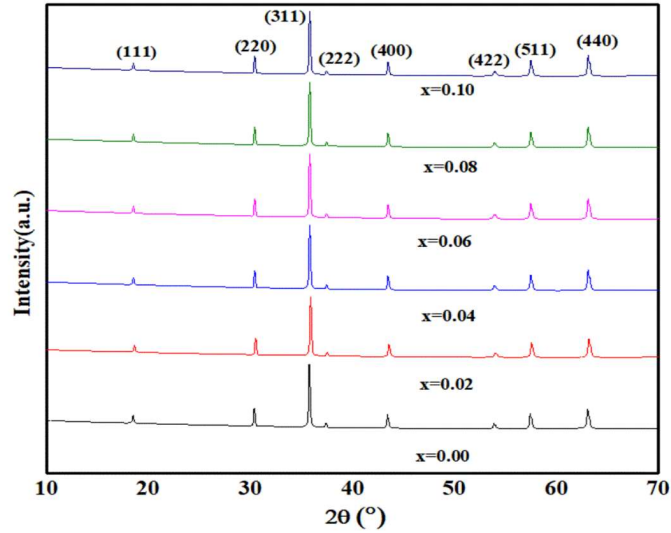

Fig. 1. XRD spectra for the synthesized $Ni_{0.7}Cu_{0.3}Al_xFe_{2-x}O_4$ ferrite nanoparticles annealed at 700°C

The sharp diffraction peaks from XRD confirms the higher crystallinity of the ferrites. The percentage of crystallinity for the prepared nanoparticles is measured by the following equation[49,50]:

$$\% \, Crystallinity = \frac{Area \; under \; the \; crystalline \; peaks}{Area \; of \; the \; all \; peaks} \times 100 \qquad (4)$$

The theoretical density ($\rho_{th}$) is calculated by the following relation [51]:

$$\rho_{th} = \frac{8M_w}{N_a a_0^3} \qquad (5)$$

where $M_w$ and $N_a$ indicate the molecular weight of the compositions and Avogadro's number, respectively. The experimental density ($\rho_{ex}$) is calculated by the following equation:

$$\rho_{ex} = \frac{M}{\pi r^2 l} \qquad (6)$$

where M, r, and $l$ represent the mass, radius, and height of the synthesized samples in tabloid shapes, respectively.

The experimental density and the theoretical density of the samples annealed at 700 °C are listed in Table 1. The porosity is found to increases as presented in Table 1, which is due to the discontinuity of the grain size, resulting in the decrease of density. The P (%) is calculated by the following relation:

$$P\,(\%) = \frac{\rho_{th} - \rho_{ex}}{\rho_{th}} \times 100\,\% \qquad (7)$$

The porosity is nothing but a relation between inter-granular and intra-granular porosity, which is shown by the following equation:

$$P\,(\%) = P_{inter} + P_{intra} \qquad (8)$$

The total displacement length per unit volume of the crystal structure can be referred to as the dislocation density ($\delta$) and the way to reduce it is to anneal the samples at high temperatures, which in turn increases their grain size [36]. This annealing is also considered as the regulator of the strength and flexibility of the crystal structure. The visible parallel lines and random lines in the crystal may indicate the displacements, which means these lines may result due to the displacement. Displacement density and particle size follow an inverse relationship with giving an error called linearity error. The dislocation density is calculated by the following equation:

$$\delta = \frac{1}{D^2} \qquad (9)$$

The length due to the deformation of an object is closely related to the pressure applied, known as the lattice strain ($\varepsilon_{ls}$). The defects caused by imperfections in the crystal structure compel atoms to deviate slightly from their normal position [52]. These structural flaws include interstitial and/or impurity atoms that cause lattice strain, which can be determined by the following relation:

$$\varepsilon_{ls} = \frac{\beta}{4\,tan\theta} \qquad (10)$$

where $\theta$ represents the angle of diffraction and $\beta$ indicates the full width at half maximum. The stacking faults are induced in the atomic planes of the crystal because of the interruption of the layered arrangement in a normal lattice structure. The stacking fault [SF] is determined by the following equation:

$$SF = \frac{2\pi^2}{45\sqrt{(3tan\theta)}} \qquad (11)$$

Various defects in the crystal structure such as displacement, plastic deformation, point defects, and domain boundary defects are considered to be the key factors of the deformation in the structure and it is assumed that this deformation occurs in one part out of nearly one million parts of the material defined as the micro strain ($\varepsilon_{ms}$). A notable feature of the micro strain is that it maximizes the peak and the following equations are introduced to comprehend it [53]:

$$\varepsilon_{ms} = \frac{\beta \cos\theta}{4} \qquad (12)$$

The ionic radii of A and B sublattices are calculated by the following relations [9,54]:

$$r_A = \sqrt{3}a_0(u - 0.25) - r_o \qquad (13)$$

$$r_B = a_0(0.625 - u) - r_o \qquad (14)$$

where $r_o$ and u represent the radius of oxygen (1.32 Å) and oxygen parameter with the value of $\frac{3}{8}$, respectively. The distance between the centers of adjacent ions is defined as the hoping length and the lengths for A-A sites, B-B sites, and A-B sites are calculated using the following equations, respectively [54,55]:

$$L_{A-A} = \frac{a_o\sqrt{3}}{4} \qquad (15)$$

$$L_{A-B} = \frac{a_o\sqrt{11}}{8} \qquad (16)$$

$$L_{B-B} = \frac{a_o}{2\sqrt{2}} \qquad (17)$$

where $a_0$ represents the lattice constant.

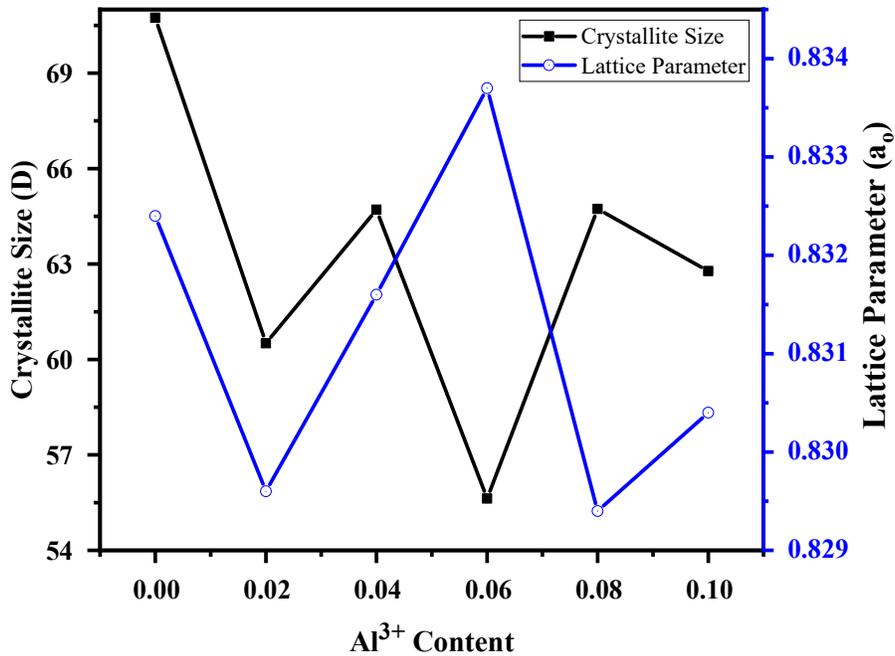

Fig. 2. Variation in lattice constant and crystallite size with $Al^{3+}$ content.

*FTIR Study*

To confirm the structure of the spinel phase in all of the prepared samples, Fourier transform infrared (FTIR) spectroscopy was utilized in this study. Fig. 3 depicts the FTIR spectra of nanocrystalline $Ni_{0.70}Cu_{0.30}Al_xFe_{2-x}O_4$ ferrites taken in the frequency region of 450-4000 cm$^{-1}$. The $v_1$ and $v_2$ are two fundamental strong absorption bands can be observed in effect of the metal-oxygen (M-O) bonds at the tetrahedral and octahedral sites. The entity of high frequency $v_1$ band found in the range of 585-615 cm$^{-1}$ which is formed by the internal stretching vibration of the M-O bonds at tetrahedral sites whereas the low frequency $v_2$ band around 400 cm$^{-1}$ corresponds to that of octahedral site [56]. The formation of spinel structures in the prepared Ni-Cu ferrite nanoparticles is ascertained by the observed bands. The bands observed in this investigation is consistent with previous findings. [57,58]. The absorption peaks, however, are induced by the tetrahedral site of the metal's intrinsic stretching vibration. Moreover, the stretching vibration of M-O at both sites is influenced by changes in the lattice parameter. The tetrahedral stretching frequency band ($v_1$) shifts to higher frequency regions as $Al^{3+}$ doping increases, as observed from the FTIR spectra. As illustrated from the Fig. 3, the observed band shifting with changing of $Al^{3+}$ concentrations might be due cations distribution followed by lighter $Al^{3+}$ substitution over the tetrahedral and octahedral sites [56,58,59].

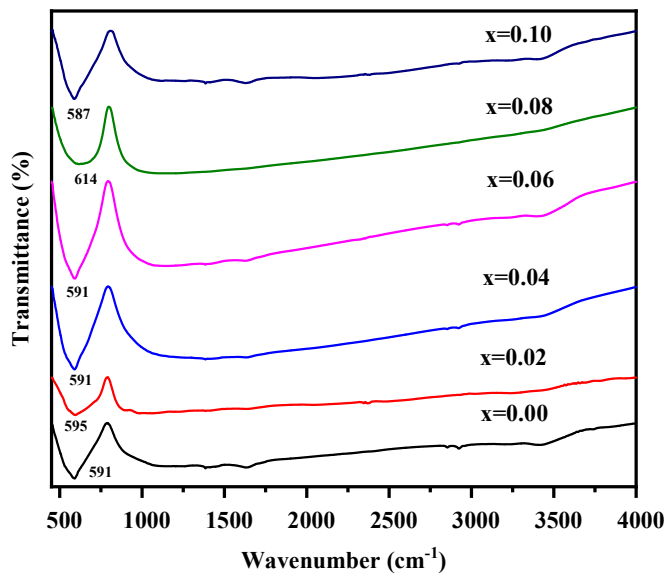

*Fig 3. Room temperature FTIR spectra of the synthesized Ni-Cu spinel ferrite nanoparticle with $Al^{3+}$ concentrations*

## 3.2 FESEM and EDX Studies

Fig. 4 shows the FESEM micrographs of nanocrystalline $Ni_{0.70}Cu_{0.30}Al_xFe_{2-x}O_4$ annealed at 700 °C. As depicted in all figures of Fig. 4 (A-F), the grains are found in semi-spherical shapes with a uniform and even distribution in multi-domains separated by grain boundaries. The average grain size of the synthesized ferrite nanoparticles is measured by [54]:

$$G_a = \frac{1.5L}{XN} \qquad (18)$$

where L, X, respectively, indicate the total length in cm and the magnification of the micrographs, and N is the number of intercepts. Fig. 5 illustrates the EDX analysis of $Ni_{0.70}Cu_{0.30}Al_xFe_{2-x}O_4$, which ensures the presence of each and every component in each sample with appropriate proportions. The expected sum of each of the observed compositions is found 100%, which confirms the accuracy of the sol-gel analysis technique and manifests its novelty.

Table 1. Structural parameters of nanocrystalline $Ni_{0.7}Cu_{0.3}Al_xFe_{2-x}O_4$ varying $Al^{3+}$ content.

| Parameters | x = 0.00 | x = 0.02 | x = 0.04 | x = 0.06 | x = 0.08 | x = 0.10 |
|---|---|---|---|---|---|---|
| d (nm) | 0.2510 | 0.2502 | 0.2508 | 0.2514 | 0.2501 | 0.2504 |
| $a_0$ (nm) | 0.8324 | 0.8296 | 0.8316 | 0.8337 | 0.8294 | 0.8304 |
| D (nm) | 70.74 | 60.51 | 64.71 | 55.63 | 64.73 | 62.78 |
| V (nm³) | 0.5768 | 0.5711 | 0.5752 | 0.5795 | 0.5705 | 0.5727 |
| % Crystallinity | 87.10 | 93.83 | 94.92 | 97.29 | 94.10 | 94.38 |
| $\delta$ (×10¹⁴) (lines/m²) | 1.998 | 2.731 | 2.338 | 3.230 | 2.387 | 2.538 |
| $\varepsilon_{LS}$ (×10⁻³) | 1.597 | 1.860 | 1.744 | 2.033 | 1.738 | 1.795 |
| SF | 0.446 | 0.445 | 0.446 | 0.446 | 0.445 | 0.445 |
| $\varepsilon_{ms}$ (×10⁻⁴) (line⁻²/m⁻⁴) | 4.9 | 5.729 | 5.357 | 6.230 | 5.355 | 5.220 |
| $\rho_{ex}$ (kg/m³) (×10³) | 3.832 | 3.832 | 3.832 | 3.832 | 3.832 | 3.832 |
| $\rho_{th}$ (kg/m³) (×10³) | 5.256 | 5.297 | 5.247 | 5.195 | 5.264 | 5.230 |
| P (%) | 27.093 | 27.657 | 26.968 | 26.237 | 27.204 | 26.730 |
| $G_a$ (nm) | 59 | 64 | 65 | 62 | 61 | 62 |
| $r_A$ (nm) | 0.0482 | 0.0476 | 0.0481 | 0.0485 | 0.0476 | 0.0478 |
| $r_B$ (nm) | 0.0761 | 0.754 | 0.0759 | 0.0764 | 0.0753 | 0.0756 |
| $L_{A-A}$ (nm) | 0.3605 | 0.3605 | 0.3605 | 0.3605 | 0.3605 | 0.3605 |
| $L_{A-B}$ (nm) | 0.3451 | 0.3451 | 0.3451 | 0.3451 | 0.3451 | 0.3451 |
| $L_{B-B}$ (nm) | 0.2943 | 0.2943 | 0.2943 | 0.2943 | 0.2943 | 0.2943 |

## 3.3 Dielectric Property

Fig.6 (A, B) demonstrates the variation in real ($\varepsilon'$) and imaginary ($\varepsilon''$) parts of complex dielectric constant of nanocrystalline $Ni_{0.70}Cu_{0.30}Al_xFe_{2-x}O_4$ annealed at 700 °C with increasing frequency. The dielectric property of ferrites is contingent on different factors such as preparation method, chemical composition, grain size, electronic di-polarity, and so on. The $\varepsilon'$, $\varepsilon''$, and dielectric loss tangent (*tan $\delta_E$*) are calculated by the following relations:

$$\varepsilon' = \frac{Ct}{\varepsilon_o A} \quad (19), \quad \varepsilon'' = \varepsilon' tan\delta_E \quad (20), \text{ and } tan\delta_E = \frac{1}{\omega\varepsilon_o\varepsilon'\rho} \quad (21)$$

where C is the capacitance, $\omega=2\pi f$, $f$ represents the applied field frequency, $\varepsilon_o$ represents the free-space permittivity, $t$ is the thickness and $A$ is the area of the contact surface of the tabloids.

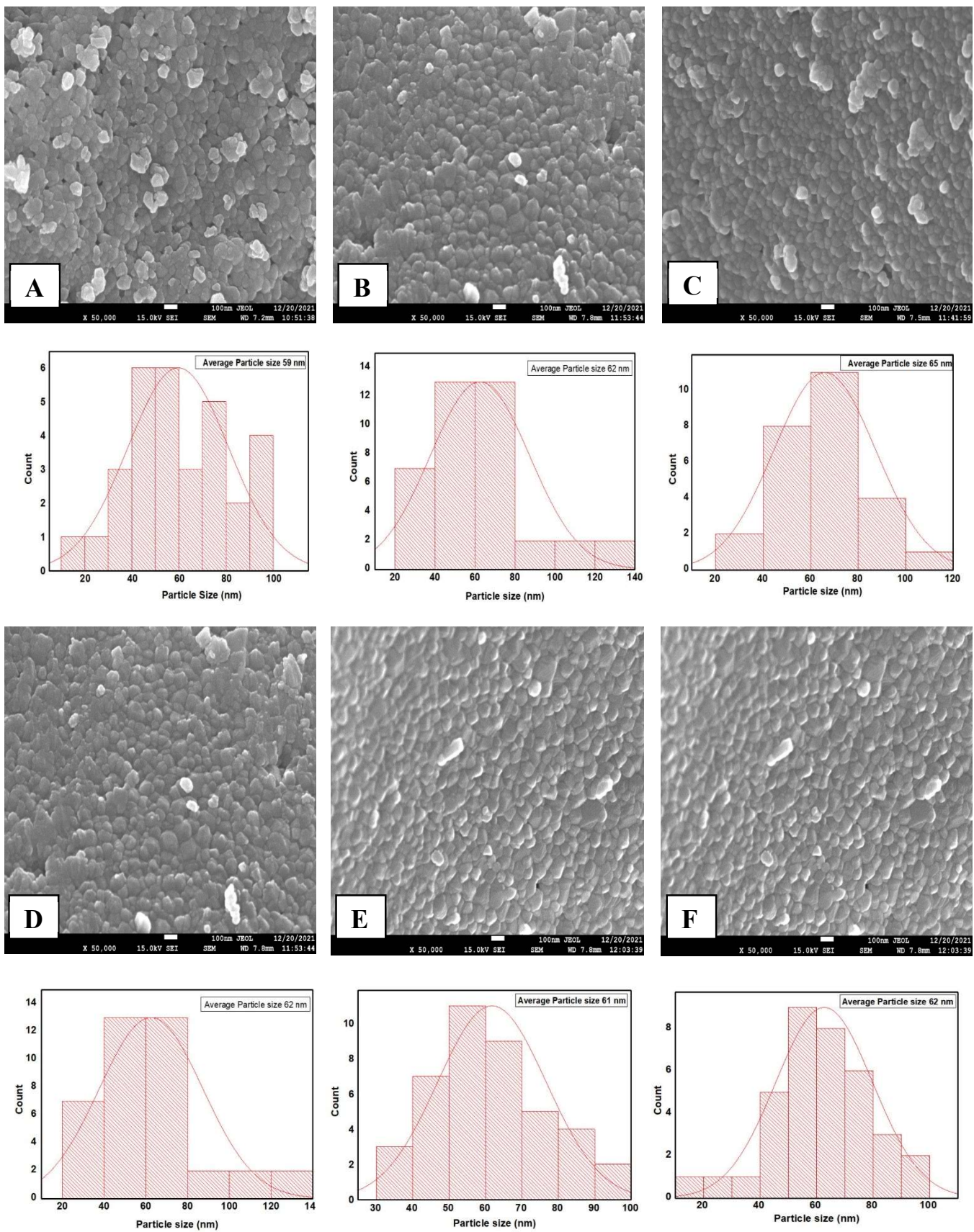

*Fig. 4. FESEM micro-graphs and corresponding histogram analysis of the synthesized $Ni_{0.7}Cu_{0.3}Al_xFe_{2-x}O_4$ NPs; ((A) x= 0.00, (B) x=0.02, (C) x=0.04, (D) x=0.06, (E) x=0.08, and (F) x=0.10)).*

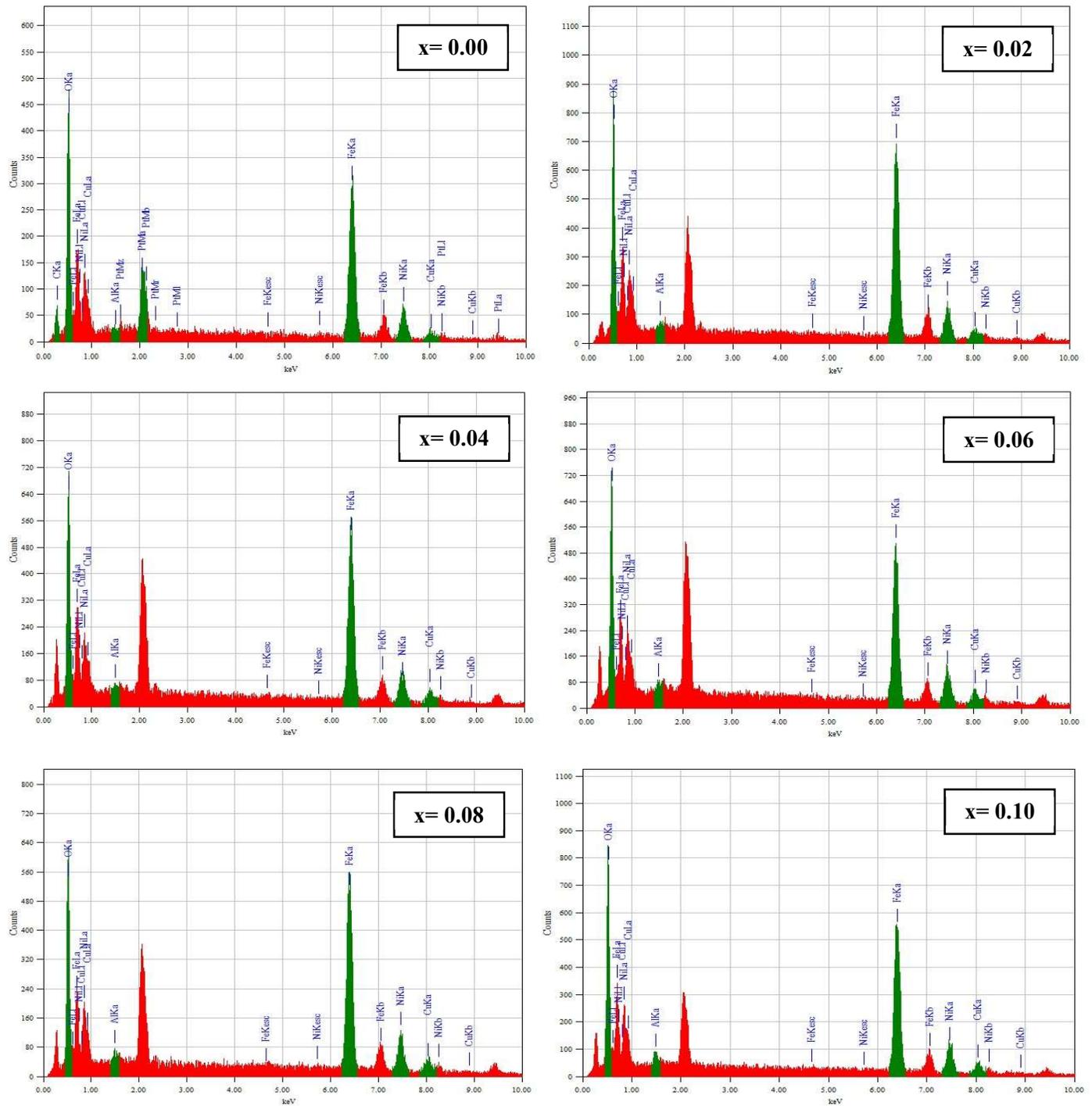

*Fig. 5. Energy dispersive spectra (EDX) of the prepared $Al^{3+}$ substituted nanocrystalline $Ni_{0.7}Cu_{0.3}Al_xFe_{2-x}O_4$.*

Fig. 6(A) illustrates that ε′ decreases with the frequency up to $10^5$ Hz and thereafter remains almost constant with showing a very low value. On the contrary, the imaginary part (ε″) reveals higher values at low frequency regime and decreases vigorously with frequency as observed in Fig. 6(B). The observed dispersive dielectric nature of these investigated materials can be described by the Maxwell–Wagner interfacial theory of polarization supported with the Koop's phenomenological theory [54,55,60]. The grain boundaries are more active at low frequencies, whereas at high frequencies grains are more contributing. At the low frequency, the value of $\varepsilon'$ is higher because of the high resistive grains, which gives the space charge polarization [47,48]. The decrease of real dielectric constant ($\varepsilon'$) with increasing frequency is found in Fig. 6A, because the grains come into action at higher frequencies and the hopping electrons cannot follow the applied electric field, causing the polarization to decrease, and the value of $\varepsilon'$ appears to be very low, becoming almost constant [61].

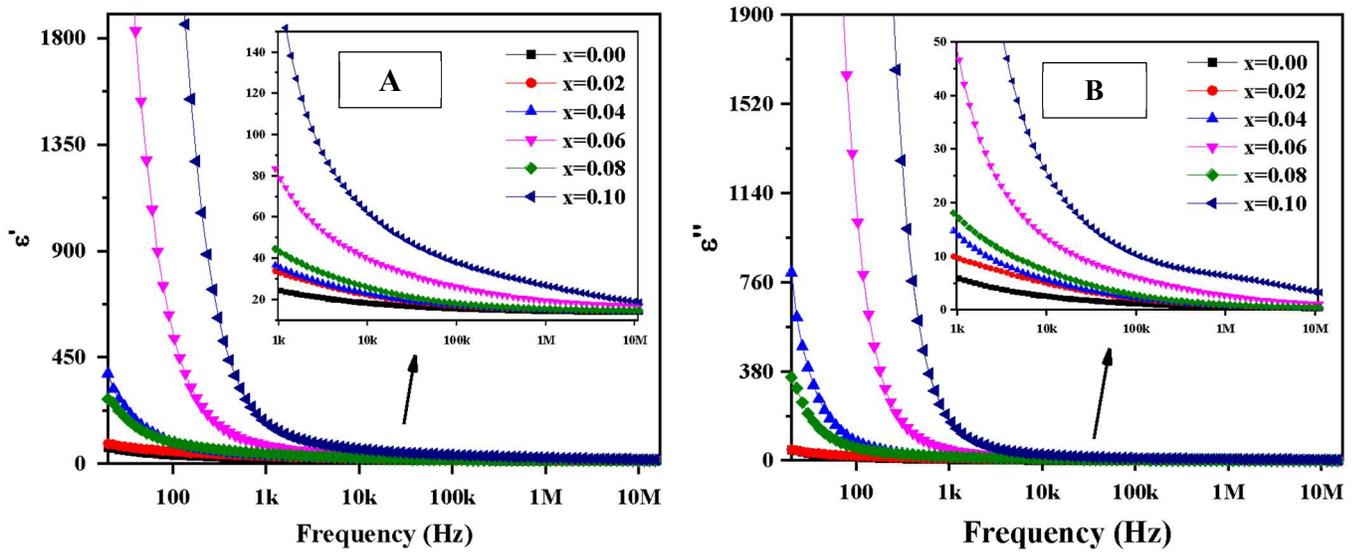

Fig. 6. Extracted initial permeability (A) real and (B) Imaginary part for the investigated nanocrystalline $Ni_{0.7}Cu_{0.3}Al_xFe_{2-x}O_4$.

The sample with x = 0.1 shows the maximum value of $\varepsilon'$ because of the redistribution of $Fe^{3+}$ at both A- and B-sites in $Ni_{0.70}Cu_{0.30}Al_xFe_{2-x}O_4$. The substitution of $Fe^{3+}$ by $Al^{3+}$ in the compositions results the transfer of $Al^{3+}$ at A-sites and replaces some $Fe^{3+}$ at B-sites, which causes the enhancement of $Fe^{3+}$ ions in the grain and assembles them in the grain boundary [48,61]. Consequently, the space charge polarization is increased and caused a higher value of the dielectric constant. Due to the generation of heat in dielectric materials by the high flow of electricity which is dissipated and considered as the material's loss that is characterized as the imaginary part ($\varepsilon''$) of dielectric constant [54]. From Fig. 6(B), it is observed that the value of $\varepsilon''$ increases significantly with increasing $Al^{3+}$ content in ferrites. The decrease of $\varepsilon''$ with frequency is occurred due to the high resistive effect of the grain boundaries. The electrons reverse their direction of motion frequently at higher frequencies and the hopping electrons can no longer follow the applied electric field, the probability of charge transport at the grain boundary decreases, resulting in the decrease of polarization, giving the low value of $\varepsilon''$ [17,55,60,61].

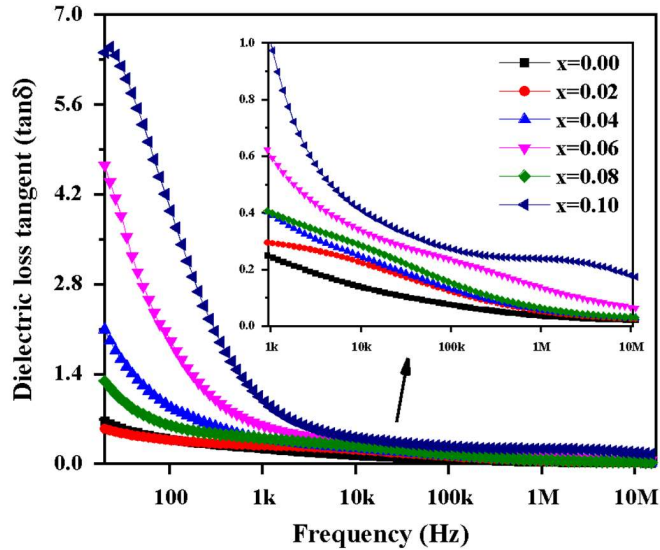

Fig. 7. Dielectric loss tangent vs frequency plot of the synthesized $Al^{3+}$ Ni-Cu NPs.

Fig. 7 shows the variation of dielectric loss tangent ($tan\ \delta_E$) of the synthesized samples annealed at 700 °C with varying frequencies. Due to impurities and imperfections, the polarization lags behind the applied voltage, causing $tan\delta_E$ to form there [54,61]. The highest value of $tan\delta_E$ is found under the relaxation condition of $\omega\tau = 1$, where $\omega = 2\pi f_{max}$, and $\tau = 1/2P$ and $f_{max}$, $\tau$ represents the peak frequency and the relaxation time, respectively and both of which are closely related to the hopping or jumping probability. Electron sharing between $Fe^{3+}$ and $Fe^{2+}$ requires very little energy and the maximum peak is achieved when the hopping frequency between them is well-matched with the applied electric field. Koop's theory explains how $tan\delta_E$ of the investigated materials decreases with frequency, in a very simple, smooth, and neat way [62,63]. It is noted that at lower conductive grain boundaries, $tan\delta_E$ exhibits the maximum value as more electrons are available to be conductive at the low-frequency region. There is energy loss that occurred during the electrons sharing between $Fe^{3+}$ and $Fe^{2+}$, therefore high energy is required [47,64,65].

The role of microstructure is important in determining the $tan\delta_E$. H. Jia et. al. showed that the grain boundaries and porosity between polycrystalline crystals affect the $\varepsilon'$ and $\varepsilon''$ [66]. The interrelation among porosity, grain boundaries, and dielectric loss is defined by the following relation:

$$tan\delta_E = (1-P)tan\delta_o + C_m P^n \qquad (22)$$

where $C_m$ is the material-dependent constant, $P$ represents the porosity and $tan\delta_o$ is the dielectric loss of material with full densification. Uniform density and lower porosity reduce the $\varepsilon'$ and $\varepsilon''$, respectively and the intrinsic and extrinsic fault are responsible for the dielectric loss.

**3.4 AC Resistivity**

The variation in ac resistivity ($\rho_{ac}$) of the investigated samples with frequencies (annealed at 700 °C) is depicted in Fig. 8. The variation in $\rho_{ac}$ of the investigated ferrite nanoparticles is explained based on the hopping mechanism. The $\rho_{ac}$ is calculated by the following equation [54]:

$$\rho_{ac} = \frac{1}{\varepsilon_0 \varepsilon' \omega tan\delta_E} \qquad (23)$$

where $\omega$ defines as the angular frequency. According to the hopping mechanism, electrons jump from one state to another, which prefer to be distributed over the sites in the lattice. In Fig. 8, it is anticipated that at lower frequencies the $\rho_{ac}$ of the investigated ferrites has higher values and depletes with increasing frequency. After a certain frequency, it gets almost saturation with showing a very small value. This variation of $\rho_{ac}$ with frequency can be described by the frequency dependency of grains and grain boundaries. The conductivity mechanism illustrates the particles ability to be highly electrically conductive [67,68].

The high-resistive boundary separating the grains are more active at lower frequencies, which impedes the movement of free charges and thus the hopping of electrons between $Fe^{2+}$ and $Fe^{3+}$ is less, which in turns result the higher values of $\rho_{ac}$ [54]. To increase the hopping of electrons between $Fe^{2+}$ and $Fe^{3+}$, it must be operated at higher frequencies, which plays a critical role in reducing the $\rho_{ac}$ value. The main reason for the low values of $\rho_{ac}$ is that the hopping of electrons almost stops after a certain frequency range. As depicted in Fig 8, the maximum value of $\rho_{ac}$ is found for the mother sample, $Ni_{0.70}Cu_{0.30}Fe_2O_4$. With the increase of $Al^{3+}$ concentration in Ni–Cu ferrites, the AC conductivity increases; therefore $Ni_{0.70}Cu_{0.30}Al_{0.1}Fe_{1.9}O_4$ shows the minimum value at the low-frequency region, and the conduction takes place through highly resistive grain boundaries, while at high frequencies conduction occurs through low resistive grains [65–67].

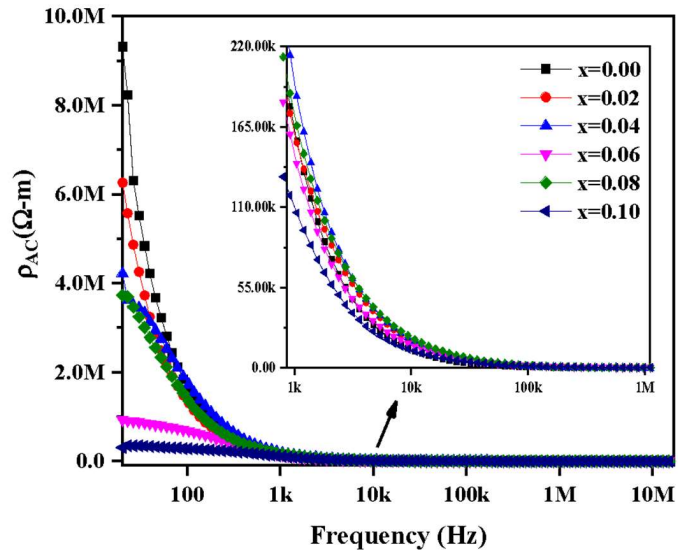

*Fig. 8. AC resistivity of the synthesized $Al^{3+}$ substituted Ni-Cu nanoparticles.*

The high-resistive boundary separating the grains are more active at lower frequencies, which impedes the movement of free charges and thus the hopping of electrons between $Fe^{2+}$ and $Fe^{3+}$ is less, which in turns result the higher values of $\rho_{ac}$[54]. To increase the hopping of electrons between $Fe^{2+}$ and $Fe^{3+}$, it must be operated at higher frequencies, which plays a critical role in reducing the $\rho_{ac}$ value. The main reason for the low values of $\rho_{ac}$ is that the hopping of electrons almost stops after a certain frequency range. As depicted in Fig 8, the maximum value of $\rho_{ac}$ is found for the mother sample, $Ni_{0.70}Cu_{0.30}Fe_2O_4$. With the increase of $Al^{3+}$ concentration in Ni–Cu ferrites, the AC conductivity increases; therefore $Ni_{0.70}Cu_{0.30}Al_{0.1}Fe_{1.9}O_4$ shows the minimum value at the low-frequency region, and the conduction takes place through highly resistive grain boundaries, while at high frequencies conduction occurs through low resistive grains [65–67].

## 3.5 Complex Electric Modulus

The electric relaxation mechanism in the materials can be explained through the spectroscopy of electric modulus (M*), which is resolved into two components [63] as given in the following:

$$M^* = \frac{1}{\varepsilon^*} = \frac{1}{\varepsilon' - i\varepsilon} = \frac{\varepsilon'}{\varepsilon'^2 + \varepsilon''^2} - i\frac{\varepsilon''}{\varepsilon'^2 + \varepsilon''^2} = M' + iM'' \qquad (24)$$

where $M' = \frac{\varepsilon'}{\varepsilon'^2 + \varepsilon''^2}$ is the real and $M'' = \frac{\varepsilon''}{\varepsilon'^2 + \varepsilon''^2}$ is the imaginary part of the electric modulus. From the above equations, both the real (M') and imaginary (M'') parts of modulus are found to be frequency-dependent, which plays a key role in investigating the relaxation mechanism of the materials. From the Fig. 9(A), it is perceived that M' responds very well to higher frequencies with exhibiting the highest value for x=0.00. It indicates the lower value of ε' at high frequencies. The inadequacy of the restorative force and the release of space charge polarization near the grain boundary helps to attain its saturation. This phenomenon occurs at higher frequencies and at the same time ensures frequency independency in the electrical properties of the materials[47,63,64].

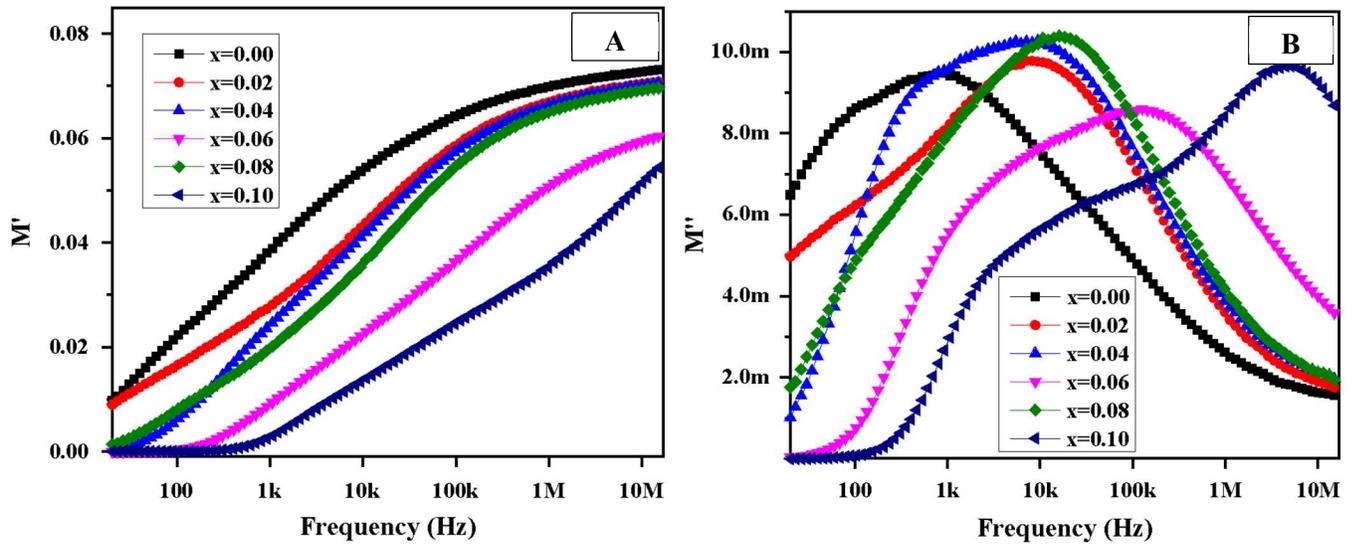

Fig. 9. Electric modulus behavior for the synthesized $Al^{3+}$ substituted Ni-Cu nanoparticles.

To illustrate the peaking behavior, one has to look at the variation of M'' as shown in Fig. 9(B). The hopping mechanism is used to illustrate the peaking behavior better as it more accurately explains the transition of the charge carriers. In the figure above, it is clear and understand that charge carriers contributing to the hopping process cover long distances at low frequencies. On the other hand, charge carriers are able to cover short distances at higher frequencies, which indicates the relaxation in the polarization process [69,70].

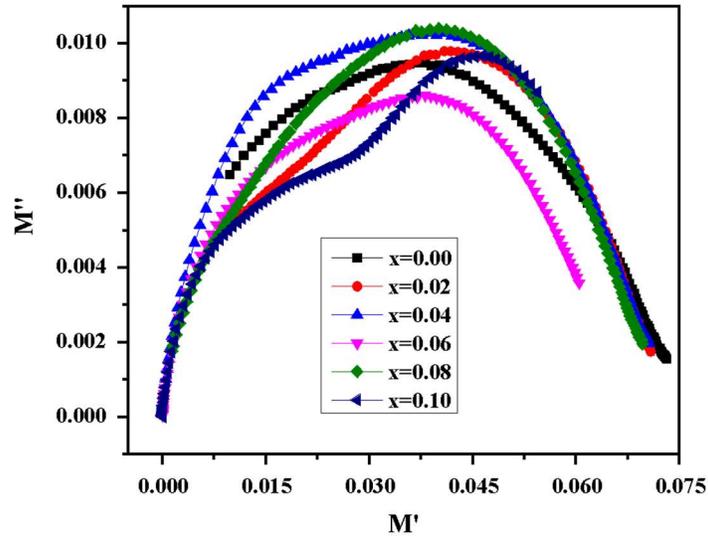

Fig. 10. M″ vs M′ plot for the investigated $Al^{3+}$ substituted Ni-Cu NPs.

The relaxation of the material is distinguished by the cole-cole plot (M″ vs M′) of the electric modulus as presented in Fig. 10. The grain and grain boundary are thought to be responsible for this separation [62,69]. A clear non-Debye type relaxation is found by looking closely at the non-overlapping semicircular pattern in Fig. 10. Nanoparticles annealed at 700 °C show two identical non- overlapping semicircular patterns [47].

3.6 Complex Impedance Analysis

To study the electrical behavior of the material, the impedance spectroscopy was employed in this study for the synthesized nanoparticles. This is a long-established method to distinguish the impedance contributions of the materials' grains, grain boundaries and electrodes. The complex impedance (Z*) includes both the resistive and reactive components of the impedance as follow:

$$Z^* = Z' - jZ'' \qquad (25)$$

where the resistive part is designated as the real part Z' which is the horizontal component of the complex impedance denoted as $Z' = |Z^*|\cos\theta$ and the imaginary part is designated as the reactive (capacitive) part expressed as $Z'' = |Z^*|\sin\theta$. However, these two components are combined impedance effect of resistance and capacitance due to grain and grain boundary which are embroiled to dielectric and electric modulus parameters following the relation:

$$\tan\delta = \frac{\varepsilon''}{\varepsilon'} = \frac{Z''}{Z'} = \frac{M''}{M'} \qquad (26)$$

The variation in real part of complex impedance (Z') of the investigated ferrites annealed at 700 °C is illustrated in Fig. 11(A) with varying frequencies. The higher values of Z' of synthesized ferrites are revealed at lower frequencies with dispersed behavior and drop sharply up to 1 KHz and thereafter it remains to constant in high frequency.

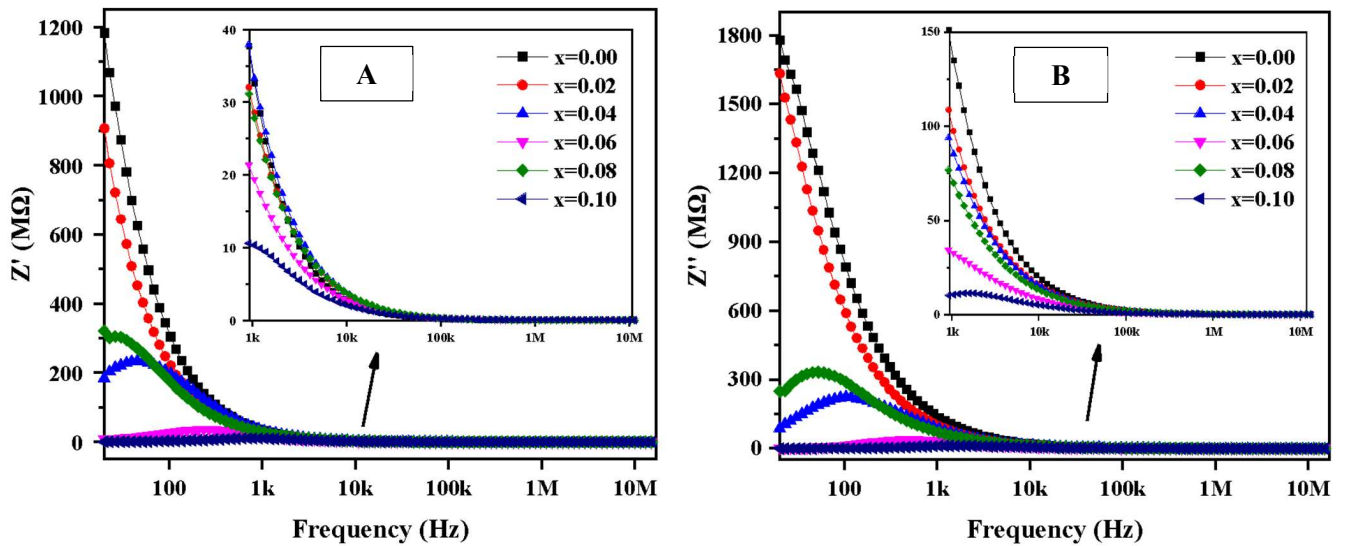

Fig. 11. Impedance analyzer extracted plots for the synthesized Al doped Ni-Cu spinel nano-ferrites.

Besides, in Fig. 11(B), the variation in imaginary part (Z'') of the complex impedance for the synthesized $Ni_{0.7}Cu_{0.3}Al_xFe_{2-x}O_4$ nano-ferrites is illustrated. As observed in Fig 11(B), the materials show higher values at the lower frequency likewise the real part (Z') and decrease rapidly with increasing frequency (up to 10 KHz) as the conductivity of the ferrites increases. However, at higher frequencies (≥100KHz), it appears with frequency-independent behavior of small constant values in effect of the reduction in polarization [54,61,71,72]. Both of the Fig. 11 shows the similar trend to the dielectric nature of the materials. For all compositions, the impedance curves are appeared to merge at higher frequencies indicating the predominance contribution of low resistive grains. Moreover, the space-charge polarization is considered important only when the materials are resolved into grains and grain boundaries [61,68]. The curves tend to converge at higher frequencies owing to a decrease in space charge polarization and this behavior elucidates the increasing tendency of ac conductivity with frequency, confirming the semiconducting behavior of the prepared nanocrystalline spinel ferrites [47,71].

The Nyquist impedance plot (also known as cole-cole plot) of the prepared $Ni_{0.7}Cu_{0.3}Al_xFe_{2-x}O_4$ ferrites annealed at 700 °C is shown in Fig. 12 which reveals the contribution of grain and grain boundary resistance as the plot is combined response of RC circuit by parallelly connected resistor and capacitor. The heterostructure nature of synthesized materials along with characteristic nature of complex impedance spectra by determining the existence of multiple electrical responses (due to grain resistance $R_g$, grain boundary resistance $R_{gb}$ and electrode effects) can be easily determined by observing the semicircular arcs appeared in the cole-cole plot [47,54,72]. By looking at Fig. 12, the two semicircular arcs are clearly visible which are formed with its center placed below the real axis, which manifests the single-phase of $Al^{3+}$ substituted nanocrystalline Ni-Cu materials. The diameter of the semicircle arcs is found to decrease with increasing $Al^{3+}$ concentration, which is actually caused by the resistance of grain boundaries. However, at lower frequencies, the $R_g$ dominates the appearance of the first semicircle, whereas at higher frequencies, the $R_{gb}$ dominates the appearance of the second semicircle. The difference in relaxation time is considered as the main catalyst behind the separation of semicircles arcs [61].

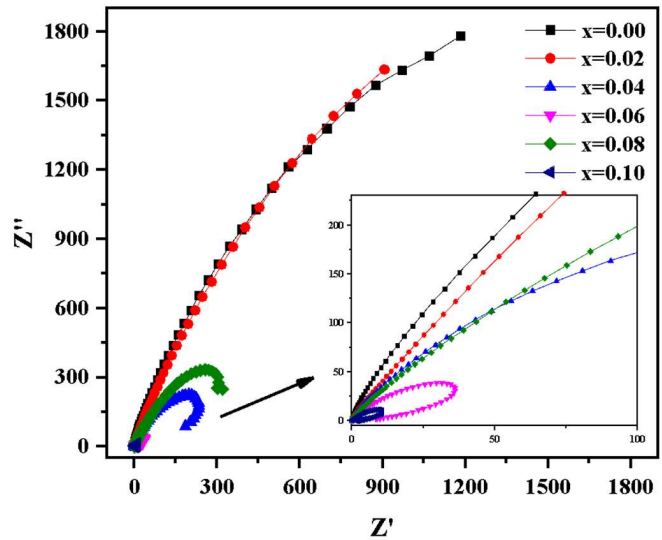

*Fig. 12. Nyquist impedance plot of the prepared Ni-Cu ferrites annealed at 700 °C*

**Conclusion**

The sol gel method was used to synthesize a series of high crystalline nanomaterials of $Ni_{0.7}Cu_{0.30}Al_xFe_{2-x}O_4$. The single-phase cubic spinel structure of the investigated materials was confirmed through XRD study with no impurity. The surface morphology was studied through the FESEM measurements, which illustrated the distribution of semi-spherical grains separated by the grain boundaries with a homogenous distribution of particles on the surface. The structural parameters were determined using the XRD and FESEM data. The electrical and dielectric properties were carried out by using the impedance analyzer supported with the modulus and impedance spectroscopy. Both the average crystallite size and the average grain size of the studied materials are found in the nano-scale range (55.63–70.74 nm) and (59.00– 65.00 nm), respectively. The dielectric dispersion nature of the materials was revealed through the dielectric study of the materials. The electrical response of the materials was inspected by means of impedance and modulus spectroscopy, which well resolved the contribution of grains and grain boundaries in the electrical properties of the investigated $Al^{3+}$ substituted Ni-Cu ferrite nanoparticles. The relaxation phenomena in the materials was justified through the cole-cole analysis of both impedance and modulus spectra. A little substitution of $Al^{3+}$ is found to be influential in the structural, dielectric and electrical properties of Ni-Cu spinel ferrites prepared by the cost-effective sol gel method.

**Acknowledgment**

The authors are grateful to the center of excellence of the Department of Mathematics and Physics at North South University (NSU), Dhaka 1229, Bangladesh. This research is funded by the NSU research grant CTRG-20/SEPS/13.

**Data Availability**

The authors are currently using all related data for the purpose of further research. If the data is requested, the authors are ready to share it with the publisher.